\documentclass[12pt]{article}

\usepackage[dvips]{graphicx}
\usepackage{amsmath}
\usepackage{amsfonts}
\usepackage{placeins}
\usepackage{afterpage}
\usepackage{dsfont}
\usepackage{flafter}     
\usepackage{scrpage}

\usepackage[a4paper]{typearea} 
  \areaset[0cm]
          {17.0cm}{27cm}           

  \newcommand{\PSImagx}[2]{\includegraphics[width=#2]{psplots/#1}}

\newcommand{\BILD}[4]{\begin{figure}[#1]%

     #2

     \centerline{\parbox{15cm}{\caption[.]{#3} \label{#4}}}
     \end{figure} }

\newcommand{\Int}{\int\limits}

\newcommand{\limacon}{lima\c{c}on }

\newcommand{\BFn}{{\boldsymbol{n}}}
\newcommand{\cA}{{\cal A}}
\newcommand{\CA}{{\cal A}}
\newcommand{\CL}{{\cal L}}

\newcommand{\BH}{{\boldsymbol{H}_1}}
\newcommand{\BHn}{{\boldsymbol{H}_1^n}}
\newcommand{\BG}{{\boldsymbol{G}}}
\newcommand{\BGo}{{\boldsymbol{G}_0}}

\providecommand{\norm}[1]{\lVert#1\rVert}

\newcommand{\R}{\mathds{R}}

\newcommand{\N}{\mathds{N}}
\newcommand{\C}{\mathds{C}}

\newcommand{\ud}{{\rm d}}
\newcommand{\ui}{{\rm i}}
\newcommand{\ue}{{\rm e}}

\newcommand{\x}{\boldsymbol{x}}	
\newcommand{\y}{\boldsymbol{y}}
	 
\newcommand{\n}{\boldsymbol{n}}  
  
\newcommand{\ta}{ {\boldsymbol{t}}}

\newcommand{\X}{\boldsymbol{\xi}}  
\newcommand{\e}{\hat {\boldsymbol{e}}}

\newcommand{\pa}{\partial}

\newcommand{\la}{\langle}
\newcommand{\ra}{\rangle}

\newcommand{\bint}{\int \limits_{\partial\Omega}}

\newcommand{\sgn}{\operatorname{sgn}}

\newcommand{\supp}{\operatorname{supp}}

\renewcommand{\Im}{\operatorname{Im}}
\renewcommand{\Re}{\operatorname{Re}}

\begin{document}


\vspace*{-1cm}

\noindent
ULM--TP/02--6

\noindent
July  2002, revised version October 2002

\newcommand{\titel}{Behaviour of boundary functions \\[1.5ex]
for quantum billiards}

\normalsize

\vspace{1.5cm}

\renewcommand{\thefootnote}{\alph{footnote}}

\begin{center}  \huge  \bf
      \titel
\end{center}

\vspace{0.5cm}

\begin{center}
   \vspace{3ex}
 
         {\large A.\ B\"acker%
\footnote{E-mail address: {\tt arnd.baecker@physik.uni-ulm.de}}$^{1)}$, 
                 S.\ F\"urstberger%
\footnote{E-mail address: {\tt silke.fuerstberger@physik.uni-ulm.de}}$^{1)}$, 
                 R.\ Schubert%
\footnote{E-mail address: {\tt schubert@spht.saclay.cea.fr}}$^{2)}$, 
          and F.\ Steiner%
\footnote{E-mail address: {\tt frank.steiner@physik.uni-ulm.de}}$^{1)}$
         }

   \vspace{4ex}

  \begin{minipage}{10.0cm}
   \begin{tabular}{ll}
   1)\hspace*{-0.25cm}&Abteilung Theoretische Physik, Universit\"at Ulm\\
   &Albert-Einstein-Allee 11, D--89069 Ulm, Germany\\[2ex]
   2)\hspace*{-0.25cm}&
   Service de Physique Th{\'e}orique, CEA/DSM/SPhT \\
   & Unit{\'e} de recherche associ{\'e}e au CNRS,
   CEA/Saclay \\
   & F-91191 Gif-sur-Yvette Cedex, France 
\\[2ex]
  \end{tabular}
  \end{minipage}

\end{center}

\renewcommand{\thefootnote}{\arabic{footnote}}
\setcounter{footnote}{0}

\vspace*{1cm}

\noindent {\bf Abstract:}

\vspace*{0.25cm}

\noindent 
We study the behaviour of the normal derivative
of eigenfunctions of the Helmholtz equation inside billiards 
with Dirichlet boundary condition. 
These boundary functions are of particular importance because they
uniquely determine the eigenfunctions inside the billiard
and also other physical quantities of interest.
Therefore they form a reduced representation of 
the quantum system, analogous to the Poincar\'e section of the classical 
system.  
For the normal derivatives we introduce 
an equivalent to the standard Green function 
and derive an integral equation on the boundary. 
Based on this integral equation 
we compute the first two terms of the mean asymptotic behaviour of the 
boundary functions for large energies. The first term is universal 
and independent of the shape of the billiard. 
The second one is proportional to 
the curvature of the boundary. 
The asymptotic behaviour is compared with numerical results
for the stadium billiard, different \limacon 
billiards and the circle billiard, and good agreement is
found.
Furthermore we derive an asymptotic completeness relation for 
the boundary functions.  

\vspace*{2cm}

\newpage

\section{Introduction}

The study of eigenfunctions of quantum systems, in particular
with chaotic classical dynamics, has attracted a lot of attention.
A prominent class of examples is provided by
billiard systems, which classically are given
by the free motion of a particle inside some domain with
elastic reflections at the boundary.
The corresponding quantum system is described by the
Helmholtz equation
inside a compact domain $\Omega \subset \R^2$ 
(in units $\hbar=1=2m$),
\begin{equation} \label{eq:Helmholtz}
  \Delta \psi_n(\x) + k^2_n \psi_n(\x) = 0 \;\; , 
            \qquad \x \in \Omega  \;\;,
\end{equation}                                                         
with (for example) Dirichlet boundary conditions 
\begin{equation} 
  \label{eq:DC-bc}
  \psi_n(\x) = 0  \;\; , \qquad \x \in \partial \Omega \;\;,
\end{equation}
where the normalized eigenfunctions $\psi_n(\x)$ are in $L^2(\Omega)$.
A detailed knowledge about the behaviour of eigenvalues
and the structure of eigenstates is relevant for applications,
for example in microwave cavities 
or mesoscopic systems (see e.g.~\cite{Sto99} and references therein).

A particular nice feature of the classical dynamics 
in  Euclidean billiards is the existence 
of a global Poincar\'e section based on the boundary $\pa \Omega$. 
This Poincar\'e section facilitates the study of the classical dynamics 
considerably, because the dynamics is reduced to an area--preserving 
map on the two--dimensional compact surface of section. 
It appears therefore desirable to look for a similar reduced representation 
of the quantum mechanical problem. One representation  
is provided by the boundary integral method. The boundary integral 
method transforms the two--dimensional Helmholtz equation \eqref{eq:Helmholtz}
for the eigenfunctions 
with Dirichlet boundary condition to a one--dimensional integral equation 
on the boundary  $\pa \Omega$.
This method involves the normal derivative of the eigenfunctions,
hereafter called the boundary functions,
\begin{equation}\label{eq:def-of-normal-der}
  u_n(s) :=  \frac{\partial}{\partial n_{\x}} \psi_n(\x)\rvert_{\x=\x(s)}  
         \equiv \la\BFn(s), \nabla \psi_n(\x(s))\ra\,\, ,
\end{equation}
where $\x(s)$ is a point on the boundary $\pa \Omega$, 
parameterized by the arclength $s$, and $\BFn(s)$ denotes the outer 
normal unit vector to $\pa \Omega$ at $\x(s)$. 
The integral equation for the boundary function is of the form 
\begin{equation}
  u=-\BH(k) u \;\;,
\end{equation}
where $\BH(k)$ is an integral operator depending on the parameter $k$ 
(for the explicit form of its integral kernel see eq.~\eqref{eq:def-of-h1} 
below). This equation has 
solutions only for a discrete set of values of the parameter $k$ which, 
when real valued, give the eigenvalues of the Helmholtz equation. The 
associated solution $u$ is the normal derivative of the corresponding 
eigenfunction which can be obtained from $u$ via an integral formula using 
the free Green function on the plane.

This reduction to the boundary is very useful for the numerical computation 
of eigenvalues and eigenfunctions. 
The boundary function allows
for a direct expression of the corresponding eigenfunction, its normalization 
\cite{Rel1940}, momentum distribution \cite{BaeSch99},
autocorrelation function \cite{BaeSch2002b} and other
quantities of interest.
Furthermore the boundary functions are the basis to define a Husimi 
representation of the eigenstates over the classical Poincar\'e section 
(see e.g.\ \cite{TuaVor95,SimVerSar97}) 
and therefore provide a direct connection with the classical 
Poincar\'e map. This is in particular useful in situations where one is 
interested in 
fine structures of the eigenstates and their relation to the classical 
dynamics, like in the field of quantum chaos. 
From this it is clear that the boundary functions
deserve a study in their own right. A profound
knowledge of their properties can then be used to obtain
a description for the above mentioned quantities.

There are two main aspects
concerning the boundary functions. First
one can consider these functions as a possible set of
basis vectors that span a kind of natural  space
for the reduced quantum system. Here the question
of orthogonality and completeness of the 
boundary functions arises.
Second, as the boundary functions contain
all information on the quantum system
inside the given domain, it is very interesting
to see how the properties of the eigenstates
of the Helmholtz equation \eqref{eq:Helmholtz}
are reflected in the boundary functions.
The last point is our main interest in this paper
where we study the mean semiclassical behaviour
of the sequence of boundary functions in terms of 
a spectral average.

A classical example of a spectral quantity is the 
spectral staircase function (integrated level density)
\begin{equation}
  N(k) := \#\{n \in \N \; | \; k_n \le k\} \;\;,
\end{equation}
whose asymptotic behaviour for $k\to\infty$ is  
given by the Weyl formula (see \cite{Hoe85b})
\begin{equation}\label{eq:Weyl_for_billiards}
  N(k) =  \frac{\CA}{4\pi} k^2
                      - \frac{\CL}{4\pi} k
                      + o(k)\;\;.
\end{equation}
Here $\CA$ denotes the area of the billiard and
$\CL$ the length of the boundary $\partial \Omega$.
It is a well known observation that the first two terms of 
\eqref{eq:Weyl_for_billiards} usually describe
the mean behaviour of $N(k)$ very well, even down to the ground state
(see for example fig.~2 in \cite{BaeSteSti95}).

In the same way one can consider the sum over a sequence of 
the normalized eigenfunctions
$\psi_n$
up to some given energy $k^2$,
\begin{equation} \label{eq:mean-efct}
  \Psi(k,\x) := \sum_{k_n\le k} \left| \psi_n(\x) \right|^2 \;\;.
\end{equation}
For billiards with $C^{\infty}$--boundary 
\cite[Theorem 17.5.10]{Hoe85a} implies
\begin{equation} \label{eq:Hoermander}
    \Psi(k,\x) 
        = \frac{1}{4\pi} k^2 - \frac{1}{4\pi} 
          \frac{J_1 \left(2d(\x) k \right)}{d(\x)} 
          k +  R(k,\x) \;\;,
\end{equation}
where $d(\x)$ is the distance of the point $\x\in\Omega$
to the boundary $\pa \Omega$, and $J_1(z)$ denotes the Bessel function. 
The remainder $R(k,\x)$ satisfies 
the estimate 
$|R(k,\x )|\leq C k$ for large $k$.
The second term in \eqref{eq:Hoermander} describes the 
influence of the boundary.
This result is, as the Weyl formula for the mean behaviour
of the spectral staircase function, completely independent
of the classical chaoticity of the underlying system.
So it applies equally well to integrable, mixed and chaotic systems.
To illustrate the behaviour of $\Psi(k,\x)$, 
let us consider a member of
the family of \limacon billiards introduced by Robnik \cite{Rob83,Rob84},
whose boundary is given in polar coordinates by 
$\rho(\varphi)=1+\varepsilon \cos\varphi$ where $\varepsilon\in[0,1]$
is the family parameter. 
In fig.~\ref{fig:mixed-systems-sum}
we show a three--dimensional 
plot of $\Psi(k,\x)/k^2$ for the desymmetrized 
\limacon billiard with parameter
 $\varepsilon=0.3$ using
the first 100 eigenfunctions of odd symmetry
(i.e. with Dirichlet boundary conditions on the symmetry axis).
For this parameter value the classical billiard has
a mixed phase space (see fig.~1 in~\cite{BaeSch2002a}).
The agreement of the asymptotic behaviour given by
\eqref{eq:Hoermander} with $\Psi(k,\x)$
is very good (see \cite{BaeSchSti98} for further examples).

\BILD{tbh}
     {
     \begin{center}
\begin{minipage}{7.4cm}
      \PSImagx{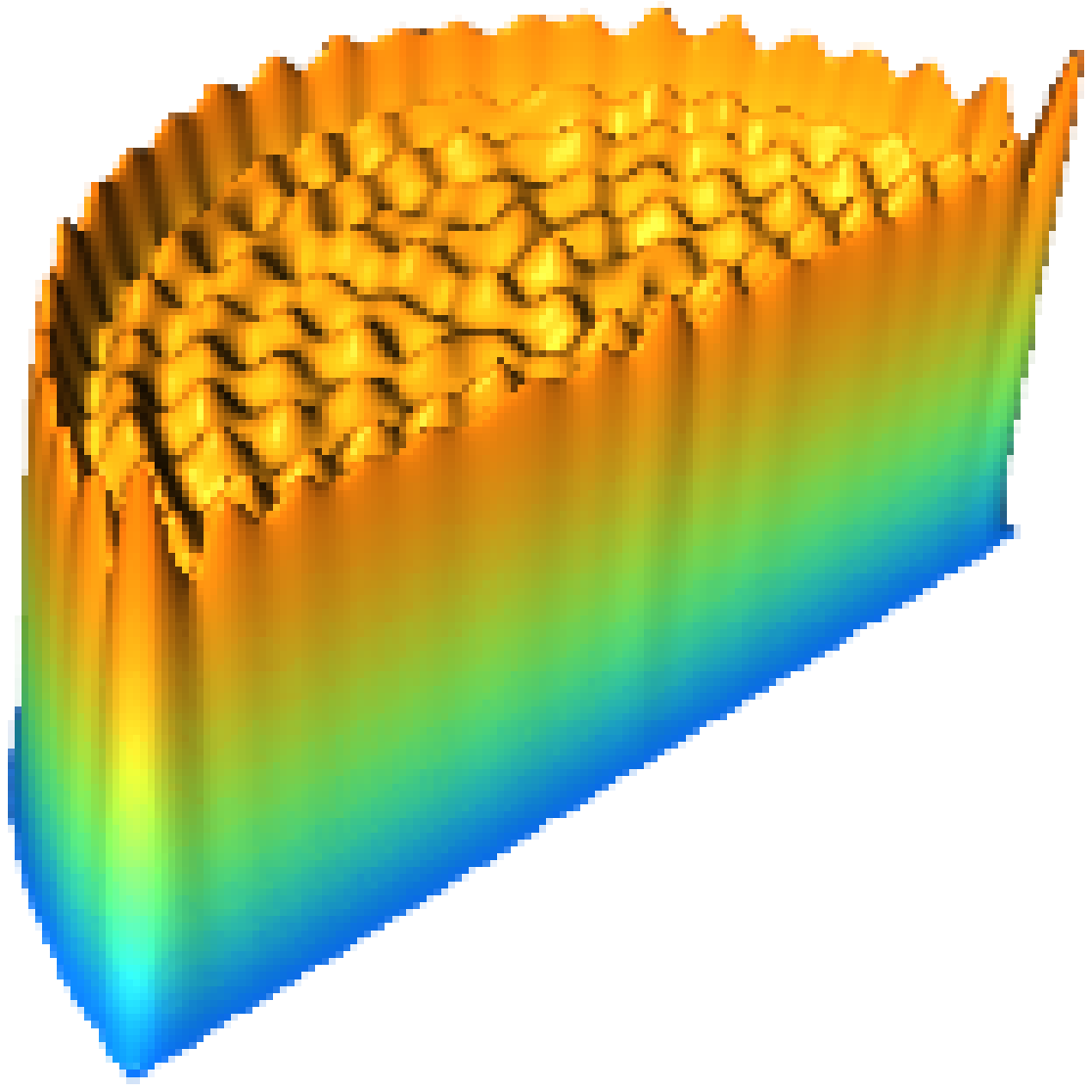}{7.1cm}
\end{minipage}
\begin{minipage}{9.4cm}
      \PSImagx{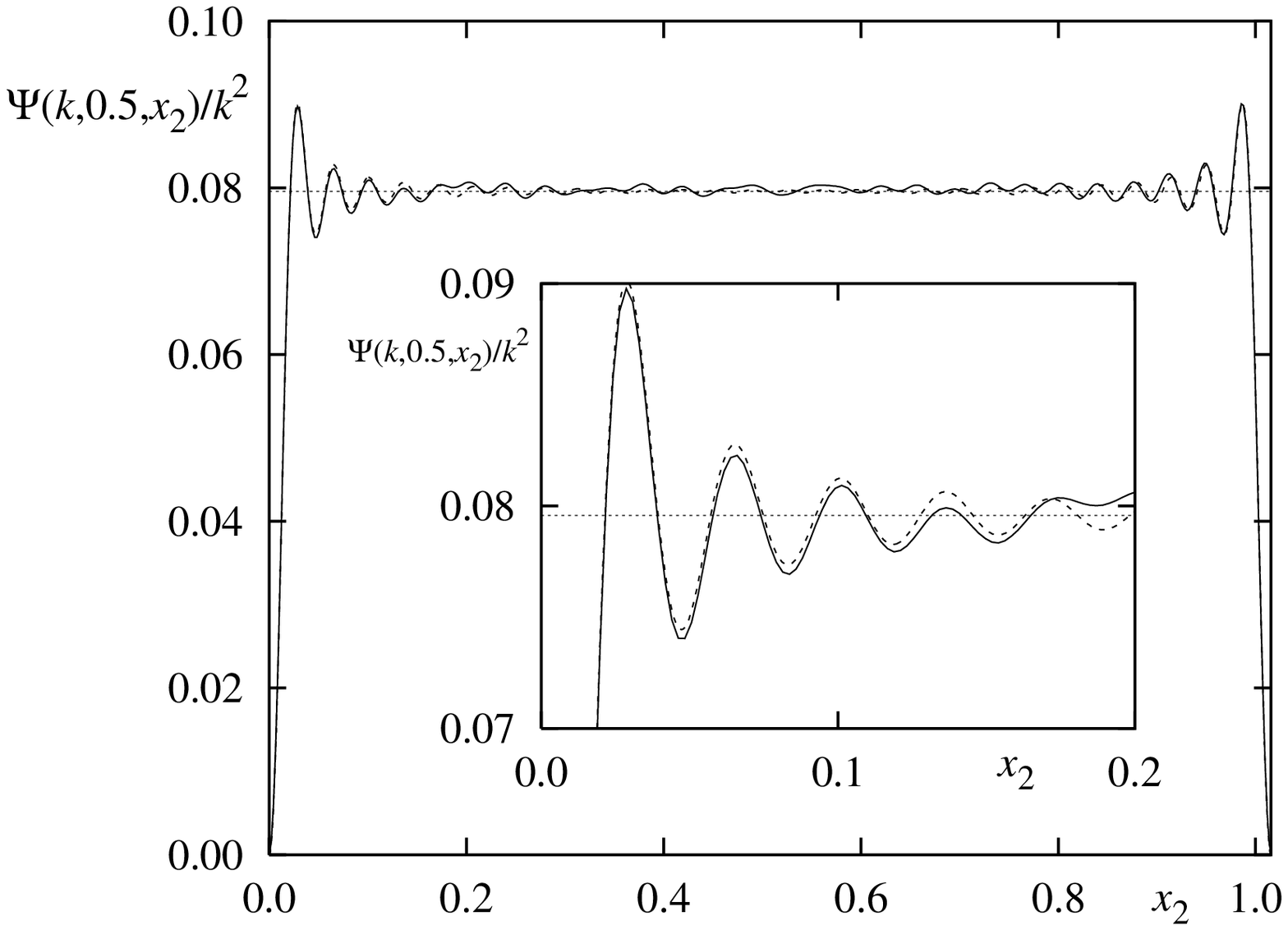}{9.3cm}
\end{minipage}
     \end{center}
     }
     {Three--dimensional plot of $\Psi(k,\x)/k^2$ for the desymmetrized 
       \limacon billiard for $\varepsilon=0.3$ and $k$ chosen
      such that $N(k)=100$. To the right a section of $\Psi(k,\x)/k^2$ 
      at $x_1=0.5$ for the case $N(k)=1000$
      is shown and compared with the asymptotic result 
      \eqref{eq:Hoermander} (dashed curve). 
      The horizontal dotted line shows the leading term $1/(4\pi)$.
      The inset shows a magnification.
     }
     {fig:mixed-systems-sum}

In analogy a similar behaviour of
a sum over the sequence of boundary functions $u_n(s)$ is expected
in the limit $k\rightarrow\infty$, i.e.\
\begin{equation} \label{eq:weyl-normal-derivative}
  \sum_{k_n\leq k} \frac{|u_n(s)|^2}{k_n^2} 
         \sim \frac{1}{4\pi} k^2 + \dots \;\;.
\end{equation}
In \cite{Oza91} it has been conjectured
that the asymptotic behaviour of 
a similar sum, $\sum_{k_n\le k} |u_n(s)|^2$, is $D_1 k^4 +D_2 \kappa(s) k^3$,
where $\kappa(s)$ is the curvature of the boundary 
at the point $s\in\partial\Omega$.
In this work we derive for the sum \eqref{eq:weyl-normal-derivative}
the asymptotic behaviour $c_1k^2+c_2\kappa(s)k$ 
including the constants.
The first constant turns out to be $c_1=\frac{1}{4\pi}$
which is consistent with the leading term of the Weyl formula 
\eqref{eq:Weyl_for_billiards}.
Multiplying \eqref{eq:weyl-normal-derivative} 
by  $\tfrac{1}{2}\la\BFn(s),\x(s)\ra$ and integrating over the 
billiard boundary gives
$N(k)$ for the left hand side, because of
the following normalization relation for $u_n(s)$ 
\cite{Rel1940}
(for alternative derivations and more general boundary conditions see 
\cite{BerWil84,Boa94})
\begin{equation}\label{eq:normalization-u}
  \frac{1}{2} \Int_{\partial \Omega} 
   \la\BFn(s),\x(s)\ra \, | u_n(s) |^2 \,\ud s = k_n^2 \;\;.
\end{equation}
For the right hand side one uses 
\begin{equation}\label{eq:integral-billiard-area} 
  \frac{1}{2} \Int_{\partial\Omega}\la\BFn(s),\x(s)\ra\; \ud s = \cA 
\end{equation}
to obtain the leading term of the Weyl formula \eqref{eq:Weyl_for_billiards}.
The next to leading term can be determined by using 
\begin{equation}\label{eq:integral-billiard-circumf}
  \Int_{\partial \Omega}  \kappa(s)
   \la\BFn(s),\x(s)\ra  \,\ud s = \CL \;\;,
\end{equation}
which for billiards with $C^\infty$--boundary
follows from $\frac{\ud}{\ud s} \ta(s)=-\kappa(s) \BFn(s)$
and partial integration 
(here $\ta(s)$ is the unit tangent vector at $s\in\partial\Omega$).

In the following we derive the full asymptotic series for
the sum in eq.~\eqref{eq:weyl-normal-derivative} 
and compute the first two terms explicitly. To this end 
in sec.~\ref{sec:integral-eq}
an integral equation on the boundary is derived which
is then used in sec.~\ref{sec:mean} to
obtain the mean behaviour of the boundary functions.
Some of the more technical details are given
in the appendices. In sec.~\ref{sec:comparison} we
provide a numerical test of the asymptotic behaviour at
finite energies.
We conclude with a short summary.

\section{An integral equation on the boundary}\label{sec:integral-eq}

Our main tool for the analysis of the semiclassical behaviour of 
the boundary functions is a
boundary Green function which we define as 
\begin{equation}\label{eq:def-greenB}
\sum_{n\in\N}\frac{1}{k^2-k_n^2} \, u_n(s)u_n^*(s')\,\, .
\end{equation}
This is the analogue of the standard Green function inside the billiard, 
where we have replaced 
the eigenfunctions by their normal derivatives on the boundary. 
Our aim in this section 
is to derive an integral equation for $\eqref{eq:def-greenB}$, which 
can be solved recursively, and therefore leads to an expansion of 
$\eqref{eq:def-greenB}$ in terms of known functions.   
The method is very similar to the one used by Balian and Bloch 
\cite{BalBlo70}, and our expansion on the boundary is the exact analogue 
of their multiple reflection expansion.

The energy--dependent Green function 
for the Dirichlet Laplacian on $\Omega$ is given by
\begin{equation}
G(E,\x,\y)=\sum_{n\in\N}\frac{\psi_n(\x)\psi_n^*(\y)}{E-k_n^2}
\end{equation}
for $(\x,\y)\in \Omega\times \Omega$, where $k_n^2$ and $\psi_n$ are the 
eigenvalues and normalized eigenfunctions of 
$\Delta$, respectively, and $E$ is a complex parameter. 
This Green function is the unique solution 
of the equation 
\begin{equation}
(\Delta_{\x} +E)G(E,\x,\y)=\delta(\x-\y)
\end{equation}
in $\Omega$ which satisfies Dirichlet boundary conditions. 
To obtain a similar relation on the boundary we define
\begin{equation}\label{eq:def-gf}
\begin{split}
g(k,s,s'):=&\frac{\pa}{\pa n_{\x}}\frac{\pa}{\pa n_{\y}} 
G(k^2,\x,\y)|_{\x=\x(s),\y=\y(s')} \\
=&\sum_{n\in\N}\frac{1}{k^2-k_n^2} u_n(s)u_n^*(s')\,\, ,
\end{split}
\end{equation}
where 
$k=\sqrt{E}\in \C$, $\Im k> 0$, is the branch of the square root of $E$ 
which has positive imaginary part.

We will now derive an integral equation for $g(k,s,s')$. Let 
\begin{equation}\label{eq:G-0}
G_0(E,\x,\y):=\frac{1}{(2\pi)^2}\Int_{\R^2} \frac{1}{E-|\X|^2}
\, \ue^{\ui \la \X, \x-\y\ra}\,\, \ud^2\xi
\end{equation}
be a free Green function which satisfies 
$(\Delta_{\x}+E)G_0(E,\x,\y)=\delta(\x-\y)$ 
on $\R^2$. This function depends holomorphically on $E$ in 
the cut plane $\C\backslash \R^+$. 
Additionally we introduce the auxiliary functions 
$f(\x,s'):=\frac{\pa}{\pa n_{\y}} G(E,\x,\y)|_{\y=\y(s')}$ 
and $f_0(\x,s'):=\frac{\pa}{\pa n_{\y}} 
G_0(E,\x,\y)|_{\y=\y(s')}$.
Then 
\begin{equation}
(\Delta_{\x}+E)(f(\x,s')-f_0(\x,s'))=0
\end{equation}
for $\x\in\Omega\setminus\pa\Omega$ and therefore we can represent 
this difference as a single layer potential 
\begin{equation}\label{eq:diff=pot}
f(\x,s')-f_0(\x,s')=SL_{\mu}(\x,s')
  :=\Int_{\pa\Omega}G_0(E,\x,\y(s)) \mu(s,s')\,\,\ud s\,\, .
\end{equation}
The density $\mu(s,s')$ is determined by the boundary condition 
\begin{equation}
-f_0(\x(s''),s')=
  \Int_{\pa\Omega}G_0(E,\x(s''),\y(s)) \mu(s,s')\,\,\ud s\,\, ,
\end{equation}
and it is a standard result from potential theory that this equation is 
solvable \cite{GueLee96}. 
We can determine $\mu(s,s')$ from the jump relations 
for a single layer potential \cite{GueLee96}
\begin{equation}\label{eq:jump-rel}
\pa^{\pm}_{n_{\x}} SL_{\mu}(\x(s),s') =
\Int_{\pa\Omega}\pa_{n_{\x}}G_0(E,\x(s),\y(s'')) \mu(s'',s')\,\,\ud s''
\pm \frac{1}{2}\mu(s,s')\,\, ,
\end{equation}
where $\pa^{\pm}_{n_{\x}}$ denotes interior (+) or exterior $(-)$ limits 
of the normal derivative, i.e., for a function $\phi(\x)$ they are  defined as 
$\pa^{\pm}_{n_{\x}} \phi(\x(s))
:= \lim_{\varepsilon\to 0} \la \BFn(s), \nabla_{\x}\, \phi (\x(s)\mp 
\varepsilon \BFn(s))\ra$
where $\BFn(s)$ denotes the outer normal unit vector to the 
boundary at $\x(s)$. 
From equation \eqref{eq:diff=pot} and the jump relations 
\eqref{eq:jump-rel} we then obtain 
\begin{equation}\label{eq:sigma=g}
g(k,s,s')=\pa^{+}_{n_{\x}}f(\x(s),s')-\pa^{-}_{n_{\x}}f(\x(s),s')
=\mu(s,s')\,\, .
\end{equation}
Applying now $\pa^{+}_{n_{\x}}$ to equation \eqref{eq:diff=pot} and using 
\eqref{eq:jump-rel} and \eqref{eq:sigma=g} leads to the desired integral 
equation on the boundary 
\begin{equation}\label{eq:int-eq}
g(k,s,s')=g_{0}(k,s,s')
-\Int_{\pa\Omega}h_1(k,s,s'')g(k,s'',s')\,\, \ud s''\,\, ,
\end{equation}
where $g(k,s,s')$ is given by \eqref{eq:def-gf} and 
\begin{align}
g_{0}(k,s,s')&=2\pa_{n_{\x}}\pa_{n_{\y}}G_0(k^2,\x(s),\y(s'))
  \,\, , \label{eq:go}\\
h_1(k,s,s')&=2\pa_{n_{\x}}G_0(k^2,\x(s),\y(s'))\,\, . \label{eq:def-of-h1}
\end{align}
Equation \eqref{eq:int-eq} is a  Fredholm  integral equation of 
second kind and can be solved 
by iteration. If we write $\BG, \BGo, \BH$ for the 
operators with integral kernels $g, g_{0}$ and $h_1$, respectively, 
the integral equation becomes
\begin{equation*}
\BG=\BGo-\BH\BG\,\, ,
\end{equation*}
which can be solved for $\BG$ as 
\begin{equation}\label{eq:operator-series}
\BG=(1+\BH)^{-1}\BGo=
\sum_{n\in\N_0}(-1)^n\BHn\BGo\,\, ,
\end{equation}
and the series converges if $\BH$ is small enough. Going back to the kernels 
of the operators this gives the expansion 
\begin{equation}\label{eq:expansion}
g(k,s,s')=g_0(k,s,s')+\sum_{n=1}^{\infty}(-1)^n
\Int_{\pa\Omega}h_n(k,s,s'')g_0(k,s'',s')\,\,\ud s''
\end{equation}
where for $n\geq 2$ we have
\begin{equation}\label{eq:hn}
h_n(k,s,s'')=
\Int_{\pa\Omega}\Int_{\pa\Omega}\cdots \Int_{\pa\Omega}
h_1(k,s,s_1)h_1(k,s_1,s_2)\cdots h_1(k,s_{n-1},s'')\,\,
\ud s_1\ud s_2\cdots\ud s_{n-1}\,\, . 
\end{equation}

Let us now discuss the convergence of the series 
\eqref{eq:operator-series}. By the integral representation 
\eqref{eq:final-h1} 
for $h_1(k,s,s')$ (derived in appendix A) we see that for 
$s\neq s'$ a positive imaginary part $\Im k$ gives an exponential damping 
factor, therefore we expect that there exists a $\gamma>0$
such that for $\Im k\geq \gamma$ the operator norm of $\BH$ satisfies 
\begin{equation}\label{eq:H1-bound}
\norm{\BH}< 1\,\, .
\end{equation}
Then, for $\Im k\geq \gamma$, the series \eqref{eq:operator-series}
would converge to a bounded operator. On the other hand, the boundary integral
method tells us that the operator $\BH$ has an eigenvalue $-1$ 
if $k^2$ is an eigenvalue of the Helmholtz equation 
\eqref{eq:Helmholtz}. 
At these values 
the series \eqref{eq:operator-series} clearly diverges, as it should, 
since $g$ has a pole there.   

It is often useful to consider spectral functions different 
from $g(k,s,s')$, and one way of introducing them such that 
the series \eqref{eq:expansion} can still be applied is based on the formula 
\begin{equation}\label{eq:deltas}
\lim_{\gamma\to 0^+}-\frac{1}{\pi}\Im \frac{2(x+\ui\gamma)}{(x+\ui\gamma)^2-k^2}
=\delta(x-k)+\delta(x+k)\,\, .
\end{equation}
Let $\rho(k)$ and $a(k)$ be  functions which are 
holomorphic in the strip $-\varepsilon< \Im k< \gamma+\varepsilon$ for 
some $\varepsilon >0$, real valued for real arguments and  $a(k)$ even. 
Furthermore assume that their product $a(k)\rho(k)$ decays faster than 
$1/k^2$ for 
$\Re k\to\pm\infty$. 
Then we define
\begin{equation}\label{eq:def-of-grho}
  g^{\rho}(k,s,s'):=-\frac{1}{ \pi }
  \Im\Int_{-\infty+\ui\gamma}^{\infty+\ui\gamma} 
2z\rho(k-z) a(z) g(z,s,s')\,\, \ud z\,\, .
\end{equation}
Under the assumption \eqref{eq:H1-bound} for $\Im k\geq \gamma$ we can insert 
the expansion \eqref{eq:expansion} and obtain a series which converges 
to the kernel of a bounded operator.  
On the other hand, 
since $g^{\rho}(k,s,s')$
is holomorphic in $\{k\in \C\,|\,\Im k>0\}$, the integral 
\eqref{eq:def-of-grho} does not depend on $\gamma$.
By taking the limit $\gamma\to 0^+$ and using 
\eqref{eq:deltas} and the definition \eqref{eq:def-gf} 
we get the representation 
\begin{equation}\label{eq:grho-expr}
g^{\rho}(k,s,s')=\sum_{n\in\N}  [\rho(k-k_n)+\rho(k+k_n)] 
   a(k_n)u_n(s)u_n^*(s')\,\, .
\end{equation}
So the role of the function $\rho$ is to select a spectral window in 
the summation, whereas  $a(k)$ acts as a weight function.

Now $\rho(k)$ and $a(k)$ can be adapted to the particular question one 
is interested in. 
In the next section we want to study the mean behaviour of the boundary 
functions, and therefore we will choose 
\begin{equation}\label{eq:rho-fourier}
\rho(k)=\frac{1}{2\pi}\Int_{\R} \hat{\rho}(t)\ue^{\ui tk}\,\, \ud t
\end{equation}
where $\hat{\rho}$
is even and  has compact support in a sufficiently small 
neighborhood of $0$. This choice obviously fulfils 
the requirements needed. 
For the weight function $a(k)$ we will choose 
\begin{equation}\label{eq:weight}
a(k)=\frac{1}{k^2+\alpha^2}
\end{equation}
with $\alpha>\gamma$. This also fulfils the requirements and 
satisfies $a(k)=1/k^2+O(1/k^4)$ for $k\to\infty$, which gives the 
correct normalization factor for the $u_n$ 
in view of \eqref{eq:normalization-u}.

\section{Mean behaviour of boundary functions} \label{sec:mean}

In this section we want to study the mean behaviour of the boundary functions 
$u_n$ for large energies $k_n^2$. We will do this by choosing a suitable 
test function $\rho$ in \eqref{eq:def-of-grho}, namely we will assume that 
there is an $\varepsilon> 0$ such that 
\begin{equation}\label{eq:cond-on-rho}
\supp\hat{\rho}\subset [-\varepsilon ,\varepsilon]\quad 
\text{and}\quad \hat{\rho}(t)=1\quad \text{for}\quad t\in 
[-\varepsilon/2, \varepsilon/2]\,\, , 
\end{equation}
and furthermore
that $\varepsilon$ is smaller than the shortest periodic orbit of the 
classical billiard flow. 
From \eqref{eq:expansion}, \eqref{eq:hn} and \eqref{eq:def-of-grho} we obtain 
the expansion 
\begin{equation}\label{eq:exp-g-rho}
g^{\rho}(k, s,s')=\sum_{n\in\N_0}g_n^{\rho}(k,s,s')
\end{equation}
with 
\begin{equation}\label{eq:g-n-rho}
g_n^{\rho}(k,s,s'):=\frac{(-1)^{n+1}}{ \pi }\;
\Im\!\Int_{-\infty+\ui\gamma}^{\infty+\ui\gamma} 2z\rho(k-z) a(z)
\Int_{\pa\Omega}h_n(z,s,s'')g_0(z,s'',s')\,\, \ud s''\,\, \ud z
\end{equation}
for $n\geq 1$, and for $n=0$ 
\begin{equation}\label{eq:g-0-rho}
g_0^{\rho}(k,s,s'):=-\frac{1}{ \pi }
\Im\!\Int_{-\infty+\ui\gamma}^{\infty+\ui\gamma} 2z\rho(k-z) a(z) 
g_0(z,s,s')\,\, \ud z\,\, .
\end{equation}
If $\rho$ is furthermore positive, the sum \eqref{eq:grho-expr} can be 
interpreted as defining a mean value of boundary functions weighted 
with the factor $a(k_n)$, where the mean is taken over a spectral window 
around $k$ defined by $\rho$. 
In the following we will assume that the billiard boundary is smooth, 
or can be obtained as the desymmetrization of a smooth billiard.
We will show in  appendix \ref{app:est-on-gn} that under the conditions 
\eqref{eq:cond-on-rho} and with $a(k)\sim 1/k^2$ for $k\to\infty$
\begin{equation}\label{eq:est-on-gn}
g^{\rho}_{n}(k,s,s)=O(k^{1-n})\,\, ,
\end{equation}
and therefore the sum \eqref{eq:exp-g-rho} provides an asymptotic expansion 
for large $k$ of the mean boundary functions. 

The explicit computation of the first terms in the expansion 
\eqref{eq:exp-g-rho} is given in appendix \ref{app:second-term}. 
Here we choose $a$ to be of the form \eqref {eq:weight}.
For $s=s'$ the first term is given by 
\begin{equation} \label{eq:first-term} 
g_0^{\rho}(k,s,s)=\frac{ k}{2\pi}+O(k^{-\infty})\,\, ,
\end{equation}
and for $s\sim s'$ one has 
\begin{equation}\label{eq:first-term-notdiag}
g^{\rho}_0(k,s,s')
=\frac{ k}{2\pi}\bigg[\frac{2}{k|s-s'|}J_1(k|s-s'|)+O(s-s')\bigg]
+O(k^{-\infty})\,\, .
\end{equation}
The second term is  for $s\sim s'$ given by 
\begin{equation}\label{eq:second-term}
g_1^{\rho}(k,s,s')=-\frac{\kappa(s)}{2\pi}\cos(2k|s-s'|)+O(1/k)\,\, ,
\end{equation}
where $\kappa(s)$ denotes the curvature of the boundary at 
$s$.

Integrating $g^{\rho}(k,s,s')$ one obtains
\begin{equation}\label{eq:smooth-counting}
\bigl(N*\rho\bigr) (k,s,s')=\Int_0^{k}g^{\rho}(k',s,s')\,\, \ud k'
\end{equation}
where $*$ denotes convolution and $N(k,s,s')$ is defined as 
\begin{equation}\label{eq:counting}
N(k,s,s'):= \sum_{k_n\leq k} \frac{u_n(s)u_n^*(s')}{k_n^2}\,\, .
\end{equation}
It is well known from the theory of spectral asymptotics that 
\begin{equation}\label{eq:tauber}
N(k,s,s')=\bigl(N*\rho\bigr) (k,s,s')(1+O(1/k))\,\, ,
\end{equation}
see \cite{DimSjo99}, and therefore from an asymptotic expansion of 
\eqref{eq:smooth-counting}
we immediately obtain the leading semiclassical behaviour of 
\eqref{eq:counting}.

Inserting \eqref{eq:exp-g-rho} and using \eqref{eq:first-term}, 
\eqref{eq:second-term} 
and \eqref{eq:est-on-gn} the asymptotic behaviour of \eqref{eq:smooth-counting}
becomes
\begin{equation}\label{eq:smooth-two-term}
  \bigl(N*\rho\bigr) (k,s,s)
    =\frac{ 1}{4\pi}\,  k^2-\frac{\kappa(s)}{2\pi}\, k+O(\ln k)\,\, ,
\end{equation}
which is the main result of this section.

Thus the leading term of \eqref{eq:tauber} reads
\begin{equation} \label{eq:Nkss-leading-term}
N (k,s,s)=\frac{ 1}{4\pi}\,  k^2 +O(k)\,\, .
\end{equation}
Assuming in addition that the set of 
$p\in[-1,1]$, such that $(p,s)$ belongs to a periodic orbit of the 
billiard map, has measure zero, we expect 
that the two--term asymptotics 
\eqref{eq:smooth-two-term} holds for $N(k,s,s)$ 
as well, but with an error term $o(k)$.
However, to prove this requires to adapt the more 
sophisticated methods from \cite{DimSjo99} or \cite{Hoe85b}, where 
similar statements are proven for $N(k)$. Note that, as discussed in 
the introduction, the relation 
between $N(k)$ and $N(k,s,s)$ is given by
\begin{equation}
N(k)=\frac{1}{2}\bint\la\n(s),\x(s)\ra N(k,s,s)\;\ud s\;\;,
\end{equation}
and therefore we recover \eqref{eq:Weyl_for_billiards} from  
\eqref{eq:smooth-two-term} using \eqref{eq:normalization-u}, 
\eqref{eq:integral-billiard-area} and \eqref{eq:integral-billiard-circumf},
up to the error term.

Often one studies billiards $\Omega$ with discrete symmetries 
and restricts the study to the corresponding
symmetry subclasses of the eigenfunctions. 
For example for a system which is symmetric with respect to reflection
at the $x_1$--axis the eigenfunctions
can be classified as either odd, fulfilling $\psi(x_1,x_2)=-\psi(x_1,-x_2)$,
or even, where  $\psi(x_1,x_2)=\psi(x_1,-x_2)$. Consequently
eigenfunctions with odd symmetry satisfy Dirichlet boundary conditions
on the symmetry axis and even eigenfunctions obey Neumann boundary conditions
on the symmetry axis.
Of course, such symmetries of $\Omega$ induce symmetries of $\pa \Omega$.
We restrict ourselves to the case of a reflection symmetry at 
a point $s_0\in\pa \Omega$.
In this case the boundary Green function \eqref{eq:def-greenB} has 
to be modified to 
\begin{equation}
g^{\pm}(k,s,s')=g(k,s_0+s,s_0+s')\pm g(k,s_0+s,s_0-s')
\end{equation}
for $s,s'$ near $s_0$ with even ($+$) or odd ($-$) symmetry. 
Using this and \eqref{eq:first-term-notdiag} we obtain for 
$s$ close to $s_0$  
\begin{equation}\label{eq:first-term-symmetry}
{g^{\rho}_0}^{\pm}(k,s,s)
=\frac{ k}{2\pi}\bigg[1\pm\frac{1}{k|s-s_0|}J_1(2k|s-s_0|)\bigg]
(1+O(1/k))\,\, ,
\end{equation}
and 
\begin{equation}
{g^{\rho}_1}^{\pm}(k,s,s)=-\frac{ \kappa(s)}{2\pi}\big[1\pm\cos(2k|s-s_0|)\big]
(1+O(1/k))\,\, .
\end{equation}
Therefore we get
\begin{equation}
(N^{\pm}*\rho) (k,s,s)=\frac{k^2}{4\pi}
\bigg[1\pm\frac{1-J_0(2k |s-s_0|)}{|s-s_0|^2 k^2}\bigg]-k 
\frac{ \kappa(s)}{2\pi}\bigg[1\pm\frac{\sin(2k|s-s_0|)}{2k|s-s_0|}\bigg]+O(\ln k)\,\, , 
\end{equation}
for $s$ close to a fixpoint $s_0$ of the symmetry.

Our results show that the mean behaviour of the normalized boundary functions 
is very similar to the mean behaviour of eigenfunctions. The crucial 
difference between the two sequences of functions $\{\psi_n\}_{n\in\N}$, 
$\{u_n\}_{n\in\N}$ is that the eigenfunctions live on a two--dimensional 
space whereas the boundary functions live on a one--dimensional space. 
Since both $u_n$ and $\psi_n$ oscillate roughly with the same 
de Broglie wave length
$2\pi/k_n$, this leads to an overcompleteness of the set $\{u_n\}_{n\in\N}$. 
This statement can be made more explicit by observing that 
\eqref{eq:first-term-notdiag} implies
\begin{equation}
  g^{\rho}(k,s,s')=\frac{2}{\pi}\, \delta(s-s')+O(1/k)\,\,. 
\end{equation}
More precisely, this means that for every 
$\varphi\in C^{\infty}(\pa\Omega)$ 
\begin{equation}\label{eq:complete-expl}
\varphi(s)= 
\sum_{n\in\N} \rho(k-k_n) \varphi_n \; u_n(s)
 +O(1/k) 
\end{equation}
holds with coefficients
\begin{equation}
\varphi_n:=\frac{\pi}{2k_n^2}\Int_{\pa\Omega}u_n^*(s')
\varphi(s')\,\, \ud s'\,\,.
\end{equation}
This follows from \eqref{eq:leading-g_0} by the method of stationary phase
(see appendix \ref{app:completeness}). 
Since $\rho$ is a rapidly decreasing function, this means that the 
boundary functions with spectral parameter $k_n$ in an interval of fixed 
width around $k$ form a complete set in the limit $k\to\infty$. 
The number of these states grows like $k$, in contrast to the number 
of all states up to energy $k^2$, which, according to the Weyl formula, 
grows like $k^2$. Therefore this result gives 
a quantitative measure of the overcompleteness of the set 
$\{u_n\}_{n\in\N}$.

\section{Numerical results for integrable, mixed and chaotic systems}
\label{sec:comparison}

\BILD{tbh}
     {
     \begin{center}
 \vspace*{-0.5cm}
      \PSImagx{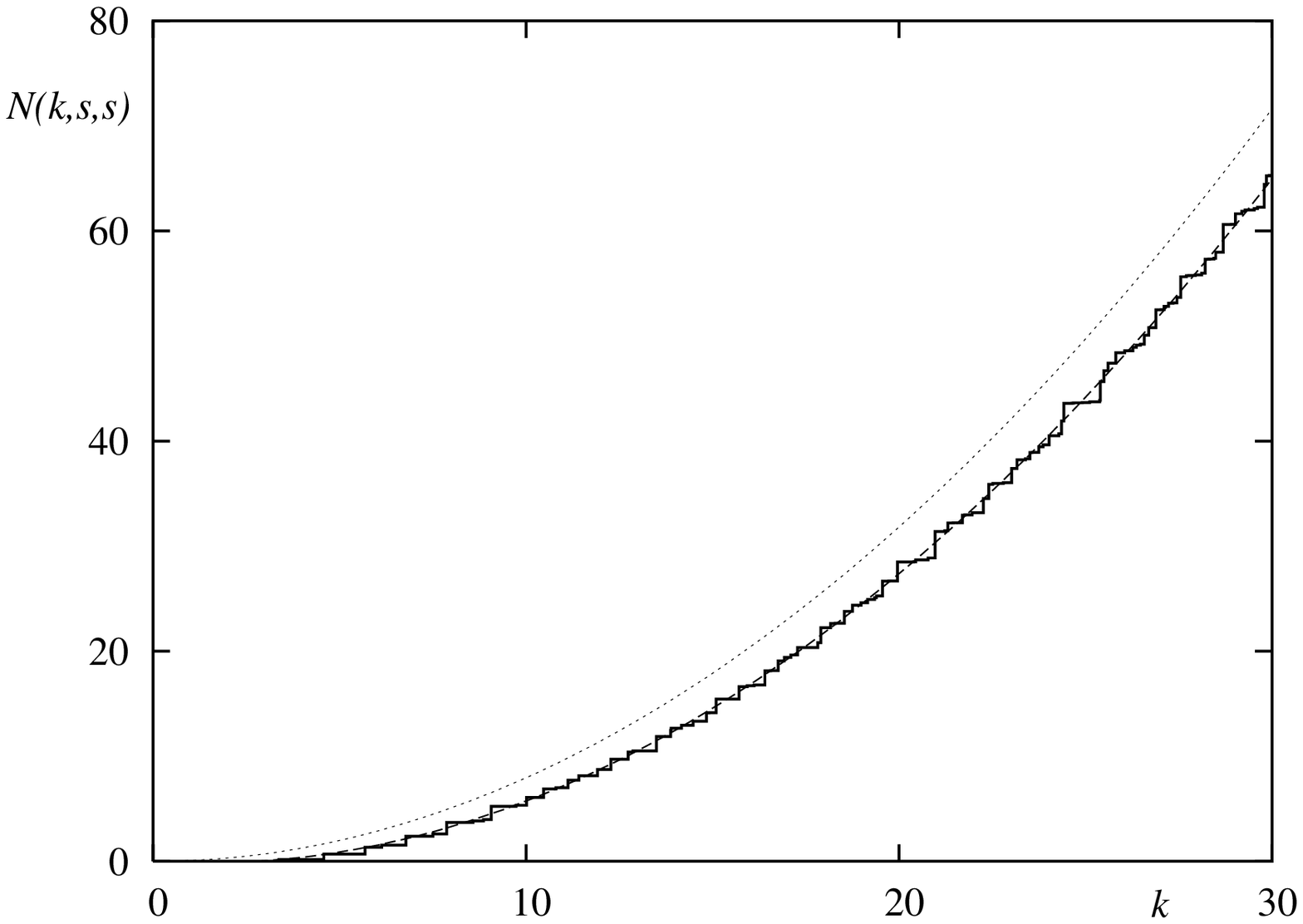}{14cm}
 \vspace*{-0.5cm}
     \end{center}
     }
     {Plot of $N(k,s,s)$ at $s=3.2$ 
      for the desymmetrized \limacon billiard with $\varepsilon=0.3$.
      The dashed curve is the asymptotic result \eqref{eq:smooth-two-term}
      and the dotted curve is just the leading term 
      \eqref{eq:Nkss-leading-term}.}
     {fig:Nkss}

In this section we would like to test how well the
asymptotic expansion \eqref{eq:smooth-two-term} 
describes the mean behaviour of the boundary function
at finite energies.
To illustrate the energy dependence we plot in fig.~\ref{fig:Nkss}
$N(k,s,s)$ for $s=3.2$ for the desymmetrized \limacon
billiard with $\varepsilon=0.3$ (see the introduction).
The dashed line beneath the staircase function (full curve)
is the asymptotic result \eqref{eq:smooth-two-term}.
As for the spectral staircase function excellent agreement,
even down to the ground state, is observed. Notice,
that for such a good agreement it is crucial to
include the curvature term. This is illustrated
by the dotted line in fig.~\ref{fig:Nkss} 
which is a plot of the leading term $k^2/(4\pi)$.

Now we turn to the $s$--dependence of $N(k,s,s)$,
where for better comparison we have divided by $k^2$ 
\begin{equation} 
  \sigma(k,s) := \frac{1}{k^2}\sum_{k_n\leq k} \frac{|u_n(s)|^2}{k_n^2} 
   \equiv \frac{1}{k^2}\; N(k,s,s) \;\;.
\end{equation}
First we consider the stadium billiard which
is given by two semi--circles joined by two parallel
straight lines.
The stadium billiard is proven to be strongly
chaotic, i.e.\ it is ergodic, mixing and a $K$-system \cite{Bun74,Bun79}.
The height of the desymmetrized billiard 
is chosen to be 1, and $a$ denotes the length
of the upper horizontal line, for which we have $a=1.8$
in the following.
The boundary is parameterized starting with $s=0$
at the corner of the quarter circle, ranging to
$s=\pi/2$ at the place where straight line and quarter circle
join tangentially until $s=\pi/2+a$ at the next corner.
Fig.~\ref{fig:stadion} shows $\sigma(k,s)$
using the first 1000 boundary functions.
The asymptotic result (see sec. \ref{sec:mean}) reads 
(for even and odd symmetry)
\begin{equation}\label{eq:stadium-with-corner}
   \overline{\sigma}^{\pm}(k,s) := 
\frac{1}{4\pi}  
\left[ 
1 \pm \sum_{\text{corners } i} \frac{1-J_0(2k|s-s_i|)}{|s-s_i|^2 k^2}
\right] -\frac{\kappa(s)}{2\pi k}\left[1\pm \sum_{\text{corners } i}
\frac{\sin(2k|s-s_i|)}{2k|s-s_i|}\right]
 \;\;, 
\end{equation}
where  the  contributions near the 
corners are taken into account.

\BILD{tbh}
     {
     \begin{center}
      \vspace*{-1cm}
      \PSImagx{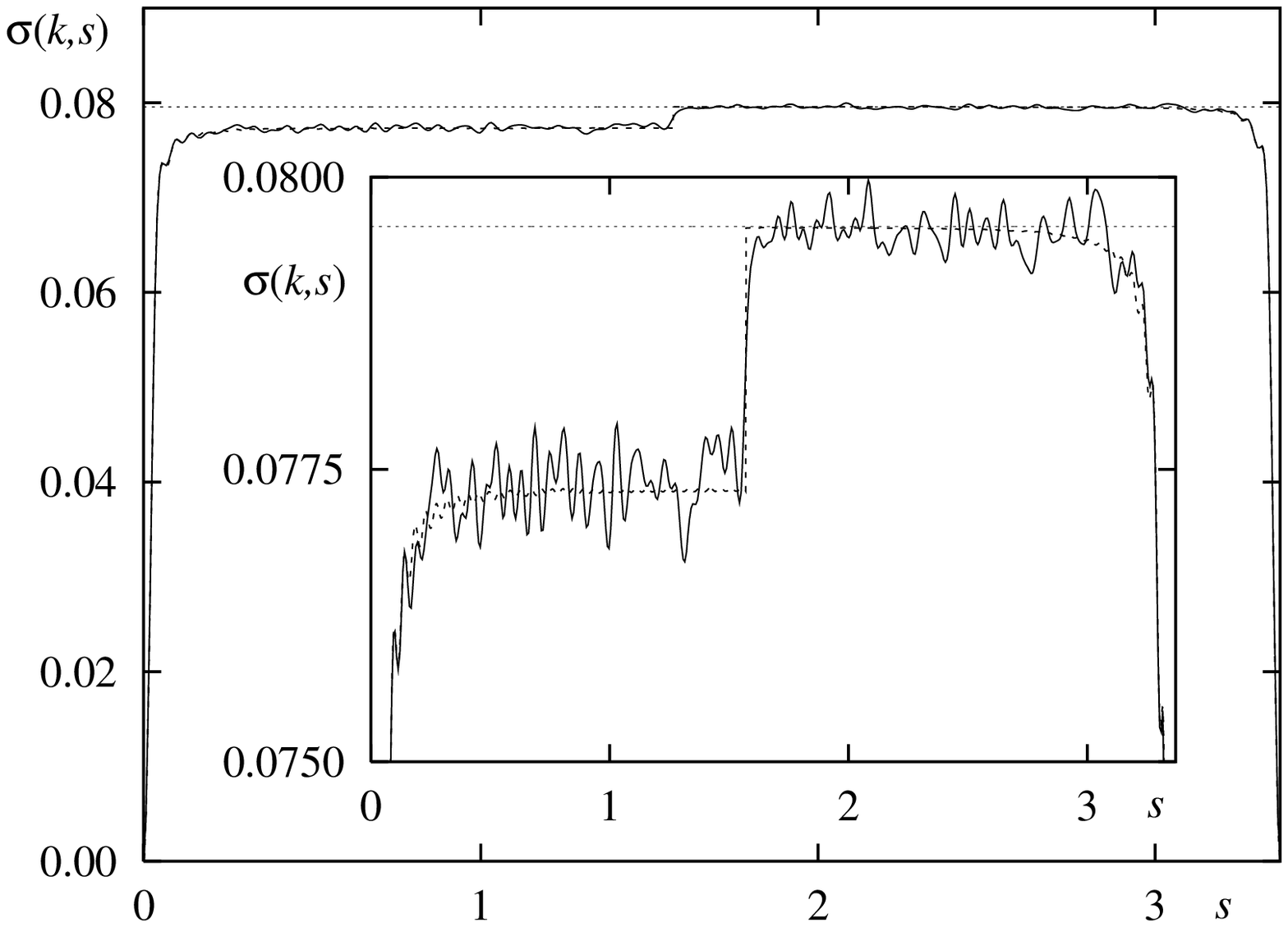}{14cm}
      \vspace*{-0.5cm}
     \end{center}
     }
     {Plot of the spectral average $\sigma(k,s)$ for 
      the fully desymmetrized
      stadium billiard with $a=1.8$ and Dirichlet boundary conditions
      everywhere. The energy $k^2$ is chosen such that the first
      1000 boundary functions are taken into account.
      The horizontal dotted line is the leading term, $1/(4\pi)$, 
      and the dashed line corresponds to  the asymptotic formula 
      $\overline\sigma^-(k,s)$, eq.~\eqref{eq:stadium-with-corner}.
      The inset shows a magnification.
     }
     {fig:stadion}

\BILD{b}
     {
     \begin{center}
      \vspace*{-1.0cm}
      \PSImagx{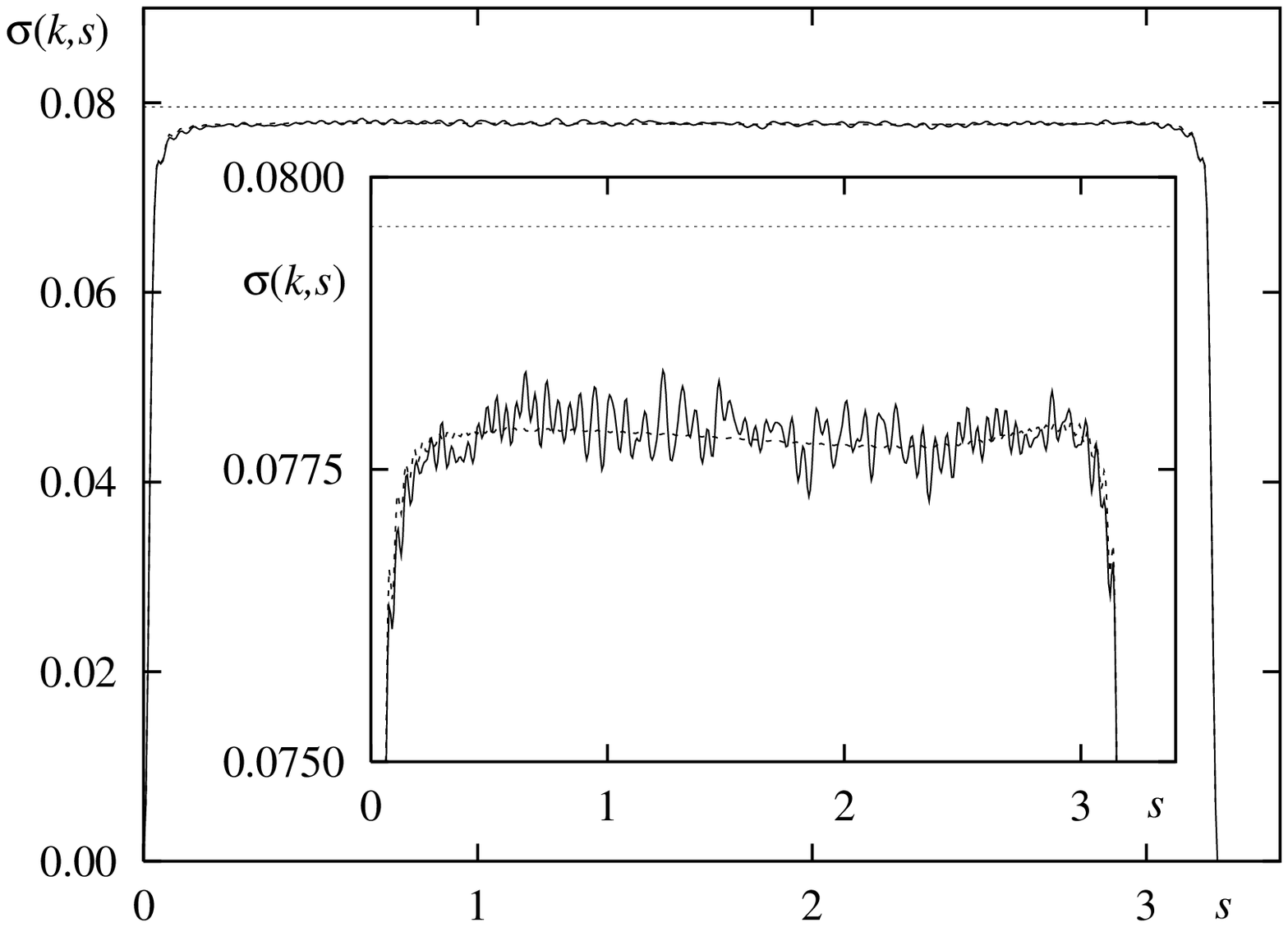}{14cm}
      \vspace*{-0.5cm}
     \end{center}
     }
     {Same as in the previous figure but for
      the desymmetrized \limacon billiard with $\varepsilon=0.3$. 
     For this parameter
     the classical phase space is mixed.
     }
     {fig:limacon-0.3}

\BILD{t}
     {
     \begin{center}
      \vspace*{-0.5cm}
      \PSImagx{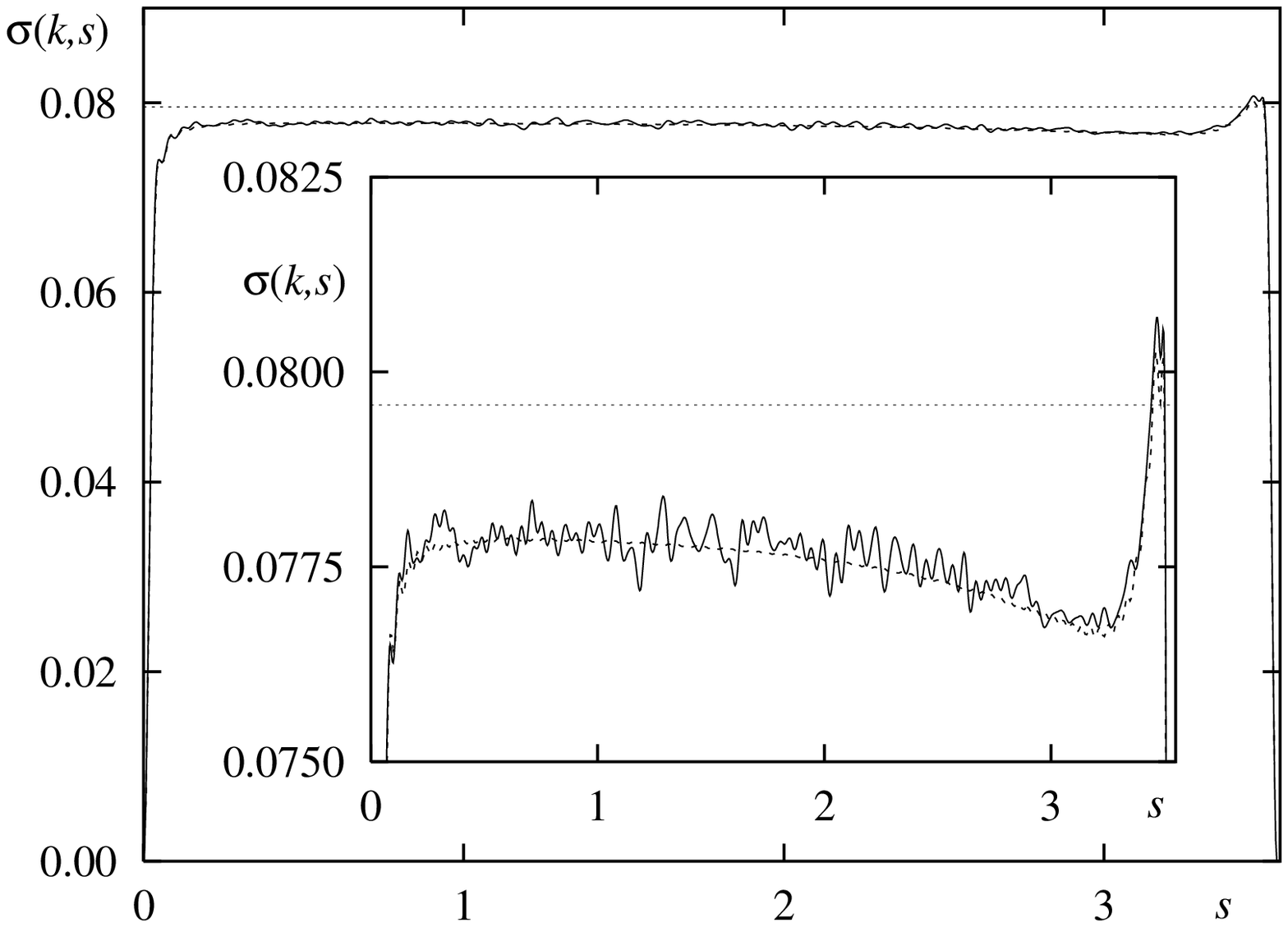}{14cm}
     \end{center}
     }
     {Same as in the previous figure but for
      the desymmetrized \limacon billiard with $\varepsilon=0.7$. 
      Here the influence of the curvature contribution
      is clearly visible.
     }
     {fig:limacon-0.7}

\BILD{b}
     {
     \begin{center}
      \vspace*{-1.0cm}
       \PSImagx{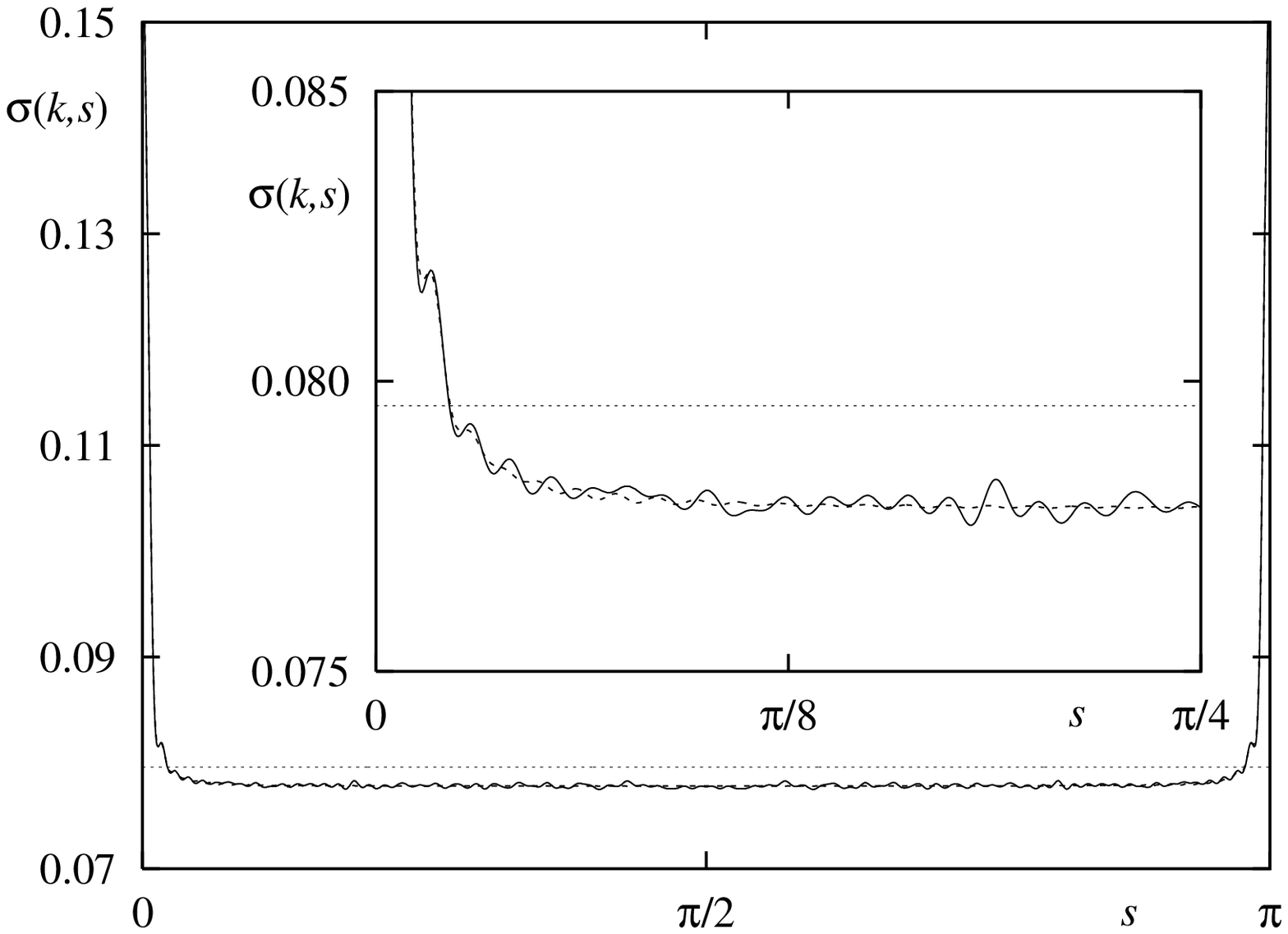}{14cm}
      \vspace*{-0.5cm}
     \end{center}
     }
     {Plot of $\sigma(k,s)$ for the desymmetrized circle billiard 
      with radius one
      and Neumann boundary condition compared with the corresponding
      asymptotic result $\overline{\sigma}^+(k,s)$ shown
      as dashed line. 
     }
     {fig:circle}

At $s=\pi/2$ the mean behaviour shows a jump
caused by the discontinuity in the curvature of
the boundary (transition from the circular part to the straight line).
Although our derivation is not valid at this point
we observe excellent agreement of ${\sigma}(k,s)$ with 
$\overline{\sigma}^{-}(k,s)$.
Moreover, the behaviour near the two corners is clearly visible
and very well described by $\overline{\sigma}^{-}(k,s)$.

The next example, shown in fig.~\ref{fig:limacon-0.3} 
for the \limacon billiard with $\varepsilon=0.3$,
illustrates that the mean behaviour is independent of the classical
dynamical properties as for this parameter the considered billiard is a
mixed system (i.e.\ regular and irregular regions in phase space
coexist).
The asymptotic formula also works well in cases where the curvature
changes more strongly as for example for the \limacon billiard with 
$\varepsilon=0.7$, see fig.~\ref{fig:limacon-0.7}.
 In this case the classical dynamics
appears to be ergodic (though it is not the case since 
there exist some very small islands of stability \cite{DulBae2001}).

Finally, we would like to consider the circle billiard as an 
example of an integrable system. 
The boundary functions of a circle billiard 
with radius $R$ and the corresponding eigenvalues are given by
\begin{equation}
u_{lm}(s)=-\frac{j_{lm}}{\sqrt{\pi}R^2}\,\ue^{\ui l\frac{s}{R}}
\quad\text{and}\quad
k_{lm}^2=\Bigl(\frac{j_{lm}}{R}\Bigr)^2 \;\;,
\end{equation}
respectively, where $j_{lm}$ denotes the $m$-th positive zero of the
$J_{l}$--Bessel function.
Therefore,
\begin{equation}
\sum_{k_{lm}\leq k}\frac{|u_{lm}(s)|^2}{k_{lm}^2}
=\frac{1}{\CA}\,N(k)\;\; .
\end{equation}
Turning to the desymmetrized circle billiard
no such simple expression exists.
Fig.~\ref{fig:circle} shows a \
plot of $\sigma(k,s)$ for the desymmetrized circle billiard
with Neumann boundary condition on the symmetry axis.
Again we find very good agreement with the corresponding
asymptotic result $\overline{\sigma}^+(k,s)$.

\section{Summary}

We have studied the semiclassical behaviour of the normal derivative
of eigenfunctions of Euclidean billiards. These boundary functions 
form a reduced representation of the quantum system, analogous 
to the Poincar\'e section of the classical billiard.
Thus they are  
of special interest in order to understand the properties of the 
eigenfunctions of billiards and their relation to the classical 
system, as investigated in the field of quantum chaos. 

We have introduced an analogue of the standard energy Green function 
for the boundary functions, and one of our main results is the derivation 
of an integral equation for this boundary Green function. The resulting
integral equation is the principal tool for the semiclassical analysis 
and leads to an expansion of the boundary Green function which is the 
analogue of the classical multiple reflection expansion. 
The resulting expansion could now be used to 
derive a semiclassical representation
of the boundary Green function in terms of orbits of 
the classical billiard map, but here we were only 
interested in the so--called length--zero contribution which determines the 
mean behaviour of the normal derivatives for large energies.  

For the mean behaviour of the sequence of the boundary functions
we have derived  a two--term asymptotics 
for large energies, 
where the first term is universal, i.e.\
completely independent of the given billiard, and the 
second term is proportional to the curvature of the boundary. 
These theoretical results fit very well with the numerical computations
for different Euclidean billiards.

Furthermore we have obtained a completeness relation for the 
boundary functions 
with momenta $k_n$ near $k$ for large $k$, which shows that the whole set 
of normal derivatives is highly over--complete, since a fraction of order $k$ 
is asymptotically sufficient to span the reduced state space 
over the boundary.

\vspace*{0.5cm}

\noindent
{\bf Acknowledgements}

\noindent
This research was supported in part by the DFG grants
Ba 1973/1--1 and Ste 241/7--3
and the EC Research Training Network under contract 
No.~HPRN--CT--2000--00103.

\FloatBarrier

\addcontentsline{toc}{section}{Appendix}
\section*{Appendix}

\begin{appendix}

\section{Integral representations}\label{app:integral_representations}
\label{app:integral-rep}

For the computation of the two leading terms in the expansion 
\eqref{eq:exp-g-rho},
eqs.~\eqref{eq:first-term} and \eqref{eq:second-term},
we need suitable integral representations of
$h_1(k,s,s')$ and $g_0(k,s,s')$.
We start by deriving a  
representation of $h_1(k,s,s')$.
Namely, from \eqref{eq:G-0} and \eqref{eq:def-of-h1} we get 
\begin{equation}
h_1(k,s,s')=\frac{2\ui }{(2\pi)^2} \Int_{\R^2}
\frac{\la \BFn(s),\X\ra}{E-|\X|^2}
\, \ue^{\ui \la \X,\x(s)-\y(s')\ra}\,\, \ud^2 \xi\,\, ,
\end{equation}
where $E=k^2$.  Inserting the relation 
\begin{equation}\label{eq:sok-plem}
\frac{1}{E-|\X|^2}=-\ui \varepsilon\Int_0^{\infty} 
\ue^{\ui \varepsilon(E-|\X|^2)t}\,\, \ud t\,\, ,
\end{equation}
with $\varepsilon =\sgn \Im E$ and for $|\Im E|>0$, allows us to solve the 
$\xi$--integral and we arrive at
\begin{equation}
h_1(k,s,s')=-\frac{\ui \varepsilon}{\pi} \Int_0^{\infty}
\frac{\la \BFn(s),\delta \x(s,s')\ra}{4t^2}
\, \ue^{\ui \varepsilon [\frac{|\delta \x(s,s')|^2}{4t}+k^2t]}\,\, \ud t
\,\, ,
\end{equation}
where we have used the abbreviation $\delta \x(s,s')=\x(s)-\y(s')$. 
This expression has the disadvantage that the factor 
$\la \BFn(s),\delta \x(s,s')\ra$ tends to zero for $s\to s'$, so we use 
\begin{equation}
-\frac{1}{ |\delta \x(s,s')|^2}\, \frac{\ud}{\ud t}
\ue^{\ui \varepsilon\frac{|\delta \x(s,s')|^2}{4t}} =
\frac{\ui\varepsilon}{4t^2}\ue^{\ui \varepsilon\frac{|\delta \x(s,s')|^2}{4t}}
\end{equation}
and partial integration (the boundary term at $t=0$ vanishes in the 
weak sense as a function of $s$) to obtain 
\begin{equation}\label{eq:final-h1}
h_1(k,s,s')=-\frac{\ui \varepsilon k^2}{\pi}
\frac{\la \BFn(s),\delta \x(s,s')\ra}{|\delta \x(s,s')|^2}
 \Int_0^{\infty}
\ue^{\ui \varepsilon[\frac{|\delta \x(s,s')|^2}{4t}+k^2t]}\,\, \ud t
\,\, .
\end{equation}
Notice that for $s \sim s'$
\begin{equation}
\frac{\la \BFn(s),\delta \x(s,s')\ra}{|\delta \x(s,s')|^2}
=\frac{1}{2}\, \kappa(s)+O(s-s')\,\,  ,
\end{equation}
where $\kappa(s)$ denotes the curvature of the boundary $\pa \Omega$ 
at $\x(s)$ (with the sign convention that it is positive for a circle).  

A similar representation can be derived for $g_{0}(k,s,s')$. 
Inserting  now 
\eqref{eq:sok-plem} in \eqref{eq:go} allows again to 
compute the $\xi$--integral and we obtain 
\begin{equation}\label{eq:pre-finalgo}
\begin{split}
g_{0}(k,s,s')=
-\frac{1}{2\pi}
 \Int_0^{\infty}\bigg[&\frac{\la \BFn(s),\delta \x(s,s')\ra
       \la \BFn(s'),\delta \x(s,s')\ra}{4r^3}\\
&\hspace*{2cm}-\ui\varepsilon\frac{\la \BFn(s), \BFn(s')\ra}{2r^2}\bigg]
\, \ue^{\ui \varepsilon[\frac{1}{4r}(\delta \x(s,s'))^2+k^2r]}
\,\,\ud r \;\;.
\end{split}
\end{equation}
The integral is again well defined as an oscillatory integral, but it will 
be useful below to have a suitable regularisation at hand when we want to 
change the order of integration. To this end, let $G_{0,\zeta}(k^2,\x,\y)$ 
be the integral kernel of the complex power of the free resolvent, 
$(-\Delta+k^2)^{-\zeta}$, $\zeta\in \C$, and let 
$g_{0,\zeta}(k,s,s'):=2\pa_{n_{\x}}\pa_{n_{\y}}G_{0,\zeta}(k^2,\x(s),\y(s'))$. 
Then a  calculation analogous to the one leading to \eqref{eq:pre-finalgo}, 
using 
$(E-|\X|^2)^{-\zeta}=\frac{\ue^{-\ui\frac{\pi}{2}\zeta}
\varepsilon^{\zeta}}{\Gamma(\zeta)}\Int_0^{\infty}
r^{\zeta-1}\ue^{\ui\varepsilon (E-|\X|^2) r}\,\, \ud r$ 
instead of \eqref{eq:sok-plem}, 
  shows that 
\begin{equation}\label{eq:pre-finalgoz}
\begin{split}
g_{0,\zeta}(k,s,s')=-\frac{\varepsilon^{\zeta+1}
   \ue^{-\ui\frac{\pi}{2}(\zeta-1)}}{2\pi\Gamma(\zeta)}
   \Int_{0}^{\infty}\bigg[&
\frac{\la\BFn(s),\delta\x(s,s')\ra
   \la\BFn(s'),\delta\x(s,s')\ra}{4r^3}\\
&\hspace*{1cm}-\ui\varepsilon
   \frac{\la\n(s),\BFn(s')\ra}{2r^2}\bigg]r^{\zeta-1}
   \ue^{\ui\varepsilon[\frac{1}{4r}(\delta\x(s,s'))^2+k^2r]}\,\,\ud r \;\;,
\end{split}
\end{equation}
and we have of course $g_{0,1}(k,s,s')=g_0(k,s,s')$.

\section{The computation of $g_0^{\rho}$ and $g_1^{\rho}$}
\label{app:second-term}

In this appendix we will use the integral representations 
derived in appendix \ref{app:integral_representations}
to compute the first and  
second term in the asymptotic expansion 
\eqref{eq:exp-g-rho}.

We start with the first term, \eqref{eq:g-0-rho}. Since the integrand is
 holomorphic in 
the upper half--plane, we can take the limit $\gamma\to 0$, and using 
\eqref{eq:go} we obtain  
\begin{equation}\label{eq:sigma-0}
g_0^{\rho}(k,s,s')
=\frac{2}{(2\pi)^2}\Int_{\R^2}[\rho(k-|\X|)+\rho(k+|\X|)] a(|\X|)
\la \BFn(s),\X\ra\la \BFn(s'),\X\ra\ue^{\ui \la \delta \x(s,s'),\X\ra}\,\, 
\ud^2 \xi \,\, .
\end{equation}
 Since $\rho$ is rapidly decreasing, and 
$|\X|$ is positive, the term $\rho(k+|\X|)$ is smaller 
than any negative power of 
$k$ for large $k$ and can therefore be neglected.  
Using the representation \eqref{eq:rho-fourier} and introducing polar
coordinates in the $\xi$--integral leads to 
\begin{equation}
\begin{split}
g_0^{\rho}(k,s,s')=\frac{2k^4}{(2\pi)^3}\Int_0^{2\pi}
\Int_0^{\infty}\Int_{-\infty}^{\infty}
\hat{\rho}(t)& a(kr)  \la \BFn(s),\e(\varphi)\ra
\la \BFn(s'),\e(\varphi)\ra\\
&\ue^{\ui k[ t(1-r)+
r\la \delta \x(s,s'),\e(\varphi)\ra]}\,\, r^3 \; \ud t\ud r\ud\varphi 
+O(k^{-\infty}) \,\, ,
\end{split}
\end{equation}
where  $\e(\varphi)$ denotes the unit 
vector in direction $\varphi$. 
The $r,t$--integrals can now be evaluated and give 
\begin{eqnarray}
&&\hspace*{-0.75cm}
\frac{k}{2\pi}\Int_{-\infty}^{\infty}\Int_{0}^{\infty}
\hat{\rho}(t)   \ue^{\ui k[ t(1-r)+
r\la \delta \x(s,s'),\e(\varphi)\ra]}\,\,a(rk) r^3 \;\ud t\ud r \\
&&=\frac{ k}{2\pi}\Int_{-\infty}^{\infty}\Int_{-1}^{\infty}
\hat{\rho}(t+\la \delta \x(s,s'),\e(\varphi)\ra)   
\ue^{-\ui k tr}\,\,a((r+1)k) (r+1) \;\ud r \ud t \;
\ue^{\ui k\la \delta \x(s,s'),\e(\varphi)\ra} \nonumber\\
&&=\ue^{\frac{\ui}{k}\pa_t\pa_r}\hat{\rho}
(t+\la \delta \x(s,s'),\e(\varphi)\ra)a((r+1)k)(1+r)|_{r=t=0} \;
\ue^{\ui k\la \delta \x(s,s'),\e(\varphi)\ra} +O(k^{-\infty}) \nonumber \\
&&=a(k)\ue^{\ui k\la \delta \x(s,s'),\e(\varphi)\ra} +O(k^{-\infty}) 
\nonumber
\end{eqnarray}
for $|\la \delta \x(s,s'),\e(\varphi)\ra|<\varepsilon/2$ by 
\eqref{eq:cond-on-rho}. In the step from the second to the third 
line we have used the general relation 
\begin{equation}
\frac{k}{2\pi}\Int_{-\infty}^{\infty}\Int_{-\infty}^{\infty}
\ue^{-\ui k rt} f(r,t)\,\, \ud r\ud t
=\ue^{-\frac{\ui}{k}\pa_r\pa_t}f(r,t)\rvert_{r=0,t=0}\,\, .
\end{equation}
The error $O(k^{-\infty})$ is due to the cut--off of the $r$--integral 
at $r=-1$.
This finally leads to 
\begin{equation}\label{eq:leading-g_0}
g_0^{\rho}(k,s,s')=\frac{a(k) k^3}{2\pi^2}\Int_0^{2\pi}
\la \BFn(s),\e(\varphi)\ra\la \BFn(s'),\e(\varphi)\ra
\ue^{\ui k\la \delta \x(s,s'),\e(\varphi)\ra}\,\, \ud\varphi 
+O(k^{-\infty})\,\, ,
\end{equation}
for $s$ close to $s'$. At $s=s'$ we get 
\begin{equation}
g_0^{\rho}(k,s,s)=\frac{a(k) k^3}{2\pi}+O(k^{-\infty})\,\, ,
\end{equation}
and for $s\sim s'$ a Taylor expansion of the exponent gives 
\begin{equation}
\begin{split}
g^{\rho}_0(k,s,s')&\approx \frac{a(k) k^3}{2\pi^2}\Int_0^{2\pi}
\cos^2\!\phi\;
\ue^{\ui k|s-s'|\sin\phi}\,\, \ud \phi\\
&=\frac{a(k) k^3}{2\pi}\frac{2}{k|s-s'|}J_1(k|s-s'|)\,\, .
\end{split}
\end{equation}

For the determination of the second term, $g_1^{\rho}(k,s,s')$, it is 
useful to use the regularized expression \eqref{eq:pre-finalgoz} with 
$\zeta$ in a range where all integrals converge, and finally make
an analytic continuation to $\zeta =1$. So we use the 
representations \eqref{eq:final-h1} and  \eqref{eq:pre-finalgoz} 
and insert them 
in \eqref{eq:g-n-rho}. 
The $k$--dependence in the expressions for $g_{0,\zeta}$ and $h_1$ 
is simple and 
we can perform the resulting $z$--integral  
\begin{equation}
\frac{1}{\pi}\Int_{-\infty+\ui\gamma}^{\infty+\ui\gamma}2z\rho(k-z)a(z)
h_1(z,s,s'')g_{0,\zeta}(z,s'',s')\,\, \ud z\,\, .
\end{equation}
Since the integrand is holomorphic in the upper half--plane, 
we can perform the 
limit $\gamma\to 0$. If we furthermore change variables 
$r\to r/k$ and $t\to t /k$ in \eqref{eq:pre-finalgo} and 
\eqref{eq:final-h1}, and interchange the order of integration, 
the $z$--integral boils down to 
\begin{equation}
\begin{split}
\frac{1}{\pi}\Int_{-\infty}^{\infty}2z^3\rho(k-z)a(z)\ue^{\ui 
\frac{t+r}{k}\, z^2}\,\, \ud z
&=\frac{1}{\pi}\,\ue^{\ui k(t+r)}\Int_{-\infty}^{\infty}2(k-z)^3a(k-z)\rho(z)
\ue^{\ui \frac{t+r}{k}\, z^2}\ue^{-\ui 2(t+r) z}\,\, \ud z\\
&=\frac{2k^3a(k)}{ \pi}\hat{\rho}(2(t+r))\ue^{\ui k(t+r)}(1+O(1/k))\,\, .
\end{split}
\end{equation}
Here we need that $a$ satisfies a kind of symbol estimate. 
Collecting the 
remaining integrals leads to 
\begin{equation}\label{eq:full-g1}
\begin{split}
g_{1,\zeta}^{\rho}(k,s,s')
&=\Im\Bigg[\frac{k^{4-\zeta}a(k)}{\pi^3}\Int_0^{\infty} \Int_0^{\infty}
\Int_{\pa \Omega}b(k,s,s',s'',r,\zeta)
\hat{\rho}(2(t+r))(1+O(1/k))\\
&\hspace*{5cm}\ue^{\ui k [\frac{1}{4r}(\delta \x(s'',s'))^2
+\frac{1}{4t}(\delta \x(s,s''))^2+(r+t)]}
\,\,\ud s''\ud r\ud t \Bigg]
\end{split}
\end{equation}
with
\begin{equation}
\begin{split}
b(k,s,s',s'',r,\zeta)
=\ue^{-\ui\frac{\pi}{2}(\zeta-1)}
\frac{\la \BFn(s),\delta \x(s,s'')\ra}{|\delta \x(s,s'')|^2}
&\bigg[\ui k\,\frac{\la \BFn(s''),\delta \x(s'',s')
\ra\la \BFn(s'),\delta \x(s'',s')\ra}{4r^3} \\
&\hspace{3cm} \;\; 
+\frac{\la \BFn(s''), \BFn(s')\ra}{2r^2}\bigg]r^{\zeta-1}\,\, .
\end{split}
\end{equation}

This looks quite complicated and to understand the properties of this 
expression better let us discuss the stationary points of the phase function 
$\frac{1}{4r}(\delta \x(s'',s'))^2+\frac{1}{4t}(\delta \x(s,s''))^2+(r+t)$
 with respect to $s'',t,r$. The derivatives with respect to $t$ and $r$ 
give the conditions 
\begin{equation}
-\frac{1}{4t^2}(\delta \x(s,s''))^2+1=0\quad ,\qquad
-\frac{1}{4r^2}(\delta \x(s'',s'))^2+1=0
\end{equation}
respectively, and hence $2t=|\delta \x(s,s'')|$ and $2r=|\delta \x(s'',s')|$. 
The $s''$--derivative leads to 
\begin{equation}
\frac{1}{2r}\langle\ta(s''),\delta \x(s'',s')\rangle-\frac{1}{2t}
\langle\ta(s''),\delta \x(s,s'')\rangle=0
\end{equation}
which yields together with the previous conditions on $t$ and $r$  
for $s''$ 
\begin{equation}
\langle\ta(s''),\widehat{\delta \x(s'',s')}\rangle=
\langle\ta(s''),\widehat{\delta \x(s,s'')}\rangle\,\, .
\end{equation}
But this is just the condition that there exists a trajectory in the billiard 
$\Omega$ starting 
at the point $s$ at the boundary, which is then elastically reflected 
at $s''$ and 
ending in $s'$. This is of course what one expects by analogy with similar 
expressions, namely that $g^{\rho}(k,s,s')$ is semiclassically given by 
a sum over all classical orbits from $s$ to $s'$, each contributing 
an amplitude depending on the stability of the orbit and an oscillating 
factor with frequency proportional to the length of the orbit. The $n$--th 
term in the expansion of  $g^{\rho}(k,s,s')$ contains exactly the orbits with 
$n$ reflections on the boundary. For the determination of the contribution 
of these orbits in leading order one can simplify the formulas for 
$g_0$ and $h_1$ considerably by using their asymptotic expansions 
for large arguments which can be easily derived by the method of stationary 
phase from the integral representations in appendix A. But if $s$ is close to 
$s'$, there is one very short orbit with $s''$ between $s$ and $s'$ whose 
length tends to zero for $s\to s'$. The contribution of this orbit determines 
the mean behaviour of $g^{\rho}(k,s,s)$ and therefore it is commonly 
called the length--zero contribution. For the computation of this 
contribution the above mentioned 
asymptotic formulas for $g_0$ and $h_1$ cannot be used because 
they are not valid in these regions, hence we must work with the 
full representation as in \eqref{eq:full-g1}.

The $s''$--integral can be solved by the method of 
stationary 
phase, and since $\hat{\rho}$ is supported in a small neighborhood of $0$
there  is  only one stationary point for $s$ close to $s'$, and we find that 
$g^{\rho}_{1,\zeta}(k,s,s')$ is for $s\sim s'$ equal to the imaginary part of 
\begin{equation}\label{eq:gafters}
\begin{split}
\frac{a(k)\kappa(s)k^{7/2-\zeta}
\ue^{-\ui\pi(\frac{\zeta}{2}-\frac{3}{4})}}{2\pi^{5/2}\,\Gamma(\zeta)}
\Int_0^{\infty}\Int_0^{\infty}  \frac{1}{r^{5/2-\zeta}}
\bigg(\frac{ t}{t+r}\bigg)^{1/2}
\ue^{\ui k[\frac{(s-s')^2}{4(t+r)}+(t+r)]}\hat{\rho}(2(t+r))\,\, 
\ud r\ud t \;\; (1+O(1/k))\,\, .
\end{split}
\end{equation}
Introducing now 
the coordinates $v=t+r$ and $w=t-r$  gives 
\begin{equation}
\begin{split}
\Int_0^{\infty}\Int_0^{\infty}  \frac{1}{r^{5/2-\zeta}}
&\bigg(\frac{ t}{t+r}\bigg)^{1/2}
\ue^{\ui k[\frac{(s-s')^2}{4(t+r)}+(t+r)]}\hat{\rho}(2(t+r))\,\, \ud r\ud t\\
& =
2^{1-\zeta} \Int_{0}^{\infty}v^{-1/2} \Int_{-v}^{v}
\frac{(v+w)^{1/2}}{(v-w)^{5/2-\zeta}}\,\, \ud w\,\, 
\ue^{\ui k[\frac{(s-s')^2}{4v}+v]}\hat{\rho}(2v)\,\, \ud v\\
&=B\bigg(\frac{3}{2},-\frac{3}{2}+\zeta\bigg)
\Int_{0}^{\infty}v^{-3/2+\zeta}\ue^{\ui k[\frac{(s-s')^2}{4v}+v]}
\hat{\rho}(2v)\,\, \ud v\\
&=B\bigg(\frac{3}{2},-\frac{3}{2}+\zeta\bigg)\\
&\hspace*{0.5cm}\times\ui \pi \hat{\rho}(0)
\left(\frac{|s-s'|}{2}\right)^{-\frac{1}{2}+\zeta}
\ue^{-\ui \pi(\frac{\zeta}{2}-\frac{1}{4})}
H^{(1)}_{\frac{1}{2}-\zeta}(k|s-s'|)[1+O(1/k)] \;\;,
\end{split}
\end{equation}
where $B(u,v)$ denotes the beta function and $H_{\nu}^{(1)}(x)$ the 
Hankel function of the first kind. 
After collecting all terms we can perform the analytic continuation to 
$\zeta =1$ and finally obtain  
\begin{equation}
g_1(k,s,s')=-\hat{\rho}(0)a(k)k^2\frac{\kappa(s)}{2\pi}\cos(k|s-s'|)[1+O(1/k)]
\end{equation}
for $s\sim s'$.

\section{Estimating $g_n^{\rho}$}
\label{app:est-on-gn}

In this appendix we derive the estimate \eqref{eq:est-on-gn} on 
$g^{\rho}_{n}(k,s,s)$. For $g^{\rho}_{n}(k,s,s)$ one obtains a similar 
expression as \eqref{eq:full-g1}, 
\begin{equation}\label{eq:full-gn}
\begin{split}
g^{\rho}_{n}(k,s,s)=\Im \Big[ k^{n+1} & \Int_0^{\infty}
\Int_{\R_{+}^n}\Int_{\pa \Omega^n}
 b(s,s',r,\tau)\tilde{a}(\tau/k)\hat{\rho}\bigg(2r+2\sum_{i=1}^n t_i\bigg)\\
&\ue^{\ui k[\frac{1}{4r}(\delta \x(s'',s'))^2+r+
\sum_{i=1}^{n+1}\frac{1}{4t_i}(\delta \x(s_{i-1}',s_i'))^2+\sum_{i=1}^n t_i]}
\,\, \ud^n s'\ud^nt\ud r\Big]
\end{split}
\end{equation}
with
\begin{equation}
\begin{split}
b(s,s',r)=\prod_{i=1}^{n+1}
\frac{\la \BFn(s_{i-1}'),\delta \x(s_{i-1}',s_i')\ra}
     {|\delta \x(s_{i-1}',s_i')|^2}
\bigg[&k\frac{\la \BFn(s_n'),\delta \x(s_n',s)
  \ra\la \BFn(s),\delta \x(s_n',s)\ra}
            {4r^3}\\
&-\ui\frac{\la \BFn(s_n'), \BFn(s)\ra}
                                     {2r^2}\bigg]\,\, ,
\end{split}
\end{equation}
where $s'=(s_1',\cdots ,s_n')$, $t=(t_1,\cdots, t_n)$ and 
we use the convention $s_0'=s_{n+1}=s$. As in the discussion after  
\eqref{eq:full-g1} one sees that the main contributions to 
\eqref{eq:full-gn} come from orbit segments, starting in $s$ 
and returning to $s$ after $n$ reflections at the boundary. 
The total length of these
orbits is $2r+2\sum_i t_i$ and since this expression appears as the argument of
$\hat{\rho}$, the only orbit contributing to 
the integral \eqref{eq:full-gn} is the one with $s_1'=s_2'=\cdots =s_n'=s$ and 
$t_i=r=0$, thanks to the small support of  $\hat{\rho}$. 
So we can approximate $(\delta \x(s_{i-1}',s_i'))^2\approx (s_{i-1}'-s_i')^2$ 
for all $i$ in the exponent of the integrand of \eqref{eq:full-gn}, 
and in the prefactor $b$ we can set $s_i'=s$ for all $i$. Now a substitution 
$s'\to s'/k$, $t\to t/k$ and $r\to r/k$ makes the integrand 
independent of $k$ up to a factor of $k^{-2n}$, which together 
with the former prefactor $k^{n+1}$  gives an overall factor $k^{-n+1}$. 
A more careful analysis shows that the resulting integrals diverge at $r=0$, 
so we should use the regularization which we already applied to the  
computation of $g_1(k,s,s')$. But this does not change the final result 
\begin{equation}
g_n(k,s,s)=O(k^{1-n})\,\, .
\end{equation}

\section{The completeness relation}\label{app:completeness}

In this appendix we derive the completeness relation 
\eqref{eq:complete-expl}. By the results of appendix \ref{app:est-on-gn}
we get 
\begin{equation}\label{eq:compl-two-term}
\Int_{\pa\Omega} \varphi(s')g^{\rho}(k,s,s')\,\, \ud s' 
=\Int_{\pa\Omega} \varphi(s')g^{\rho}_0(k,s,s')\,\, \ud s'
+\Int_{\pa\Omega} \varphi(s')g^{\rho}_1(k,s,s')\,\, \ud s'
+O(1/k)
\end{equation}
if $\varphi\in C^{\infty}(\pa\Omega)$ and 
$\rho$ satisfies the conditions \eqref{eq:cond-on-rho}. 
The first term on the right hand side can be easily computed by using 
\eqref{eq:leading-g_0}, the method of stationary phase, and by observing 
that due to the cutoff introduced by $\hat{\rho}$ the only stationary points 
come from $s'=s$. The result is 
\begin{equation}
\Int_{\pa\Omega} \varphi(s')g^{\rho}_0(k,s,s')\,\, \ud s'
=\frac{2}{\pi}\, \varphi(s)+O(1/k)\,\, ,
\end{equation}
where we have assumed that $a(z)=1/z^2+O(1/z^4)$, see \eqref{eq:weight}. 
The computation of the second 
term in \eqref{eq:compl-two-term} is similar but more complicated. 
Using now \eqref{eq:full-g1} and solving the resulting 
$s'$ and $s''$--integrals with the method of stationary phase leads to 
\begin{equation}\label{eq:step-second}
\begin{split}
\Int_{\pa\Omega} & \varphi(s')g^{\rho}_{1,\zeta}(k,s,s')\,\, \ud s' \\
&=\Im\bigg[ Ck^{3-\zeta}\kappa(s)\varphi(s)\ue^{-\ui\frac{\pi}{2}(\zeta+1)} \Int_0^{\infty}\Int_0^{\infty}
\frac{t^{1/2}}{r^{5/2-\zeta}}
 \hat{\rho}(2(t+r))
\ue^{\ui k(t+r)}\,\,\ud r\ud t\bigg](1+O(1/k))\,\, ,
\end{split}
\end{equation}
where we have collected all factors not depending on $k$ and $s$ in the real 
constant $C$.
Introducing new coordinates $v=t+r$, $w=t-r$ allows to solve the integrals, 
and after setting $\zeta =1$ we arrive at   
\begin{equation}
\Int_{\pa\Omega} \varphi(s')g^{\rho}_1(k,s,s')\,\, \ud s'
=C'' \frac{\kappa(s)}{k}\, \varphi(s)+O(1/k^2)=O(1/k)
\end{equation}
for $k\to\infty$, with another constant $C''$. Therefore the result 
\eqref{eq:complete-expl} is established.

\end{appendix}


\end{document}